\documentclass[11pt,a4paper,oneside,onecolumn]{article}
\usepackage[]{fontenc}
\usepackage[dvips]{epsfig}

\makeatletter
\def\LyX{L\kern-.1667em\lower.25em\hbox{Y}\kern-.125emX\spacefactor1000}%
\newcommand{\lyxtitle}[1] {\thispagestyle{empty}
\global\@topnum\z@
\section*{\LARGE \centering \sffamily \bfseries \protect#1 }
}

\newcommand{\lyxletterstyle}{
\setlength\parskip{0.7em}
\setlength\parindent{0pt}
}

\makeatother

\lyxletterstyle
\begin{document}

\vspace*{\fill}
\lyxtitle{\large \protect Broadcasting of Entanglement and Universal Quantum
 Cloners\LARGE   }
\vspace*{\fill}

{\large \hfill{}\normalsize Somshubhro Bandyopadhyay\( ^{a,} \)\footnote{
email: dhom@boseinst.ernet.in
} and Guruprasad Kar\( ^{b} \) \hfill{} \large \par}

{\em \( ^{a} \)Department of Physics, Bose Institute, 93/1 A.P.C. Road, Calcutta
-700009, India\hfill{}\par}

{\em \( ^{b} \)Physics and Applied Mathematics Unit, Indian Statistical Institute,
203 B.T. Road, Calcutta -700035, India\par}

\begin{abstract}

We study broadcasting of entanglement where we use universal quantum
cloners (in general less optimal) to perform local cloning operations
. We show that there is a lower bound on the fidelity of the universal
quantum cloners that can be used for broadcasting. We prove that an
entanglement is optimally broadcast only when optimal quantum cloners
are used for local copying. We also show that broadcasting of entanglement
into more than two entangled pairs is forbidden using only local operations.

\end{abstract}

\section{\large Introduction}

Broadcasting quantum inseparability i.e. nonlocal correlations of
quantum states  was first shown to be possible by Buzek et al \( \left[ 1\right]  \). The
entanglement originally shared by a single pair is transferred into
two less entangled pairs using only local operations. Suppose two
distant parties \( a_{1} \) and \( a_{2} \) share an entangled two qubit state
\begin{equation}
\label{1}
\left| \psi \right\rangle =\alpha \left| 00\right\rangle _{a_{1}a_{2}}+\beta \left| 11\right\rangle _{a_{1}a_{2}}
\end{equation}

where \( \alpha ^{2}+\beta ^{2}=1 \) and \( \alpha ,\beta  \) are real. \\
The first qubit belongs to \( a_{1} \) and the second belongs to \( a_{2} \). Each of
the two parties now performs local cloning operations on their own
qubit. It turns out that for some values of \( \alpha  \),

\( \left( a\right)  \) non local output states are inseparable, and

\( \left( b\right)  \) local output states are separable

hold simultaneously. Buzek et al. [1] used optimal quantum cloners
[3,4,5,6] for local copying of the subsystems and showed that the
nonlocal outputs are inseparable if

\begin{equation}
\label{2}
\frac{1}{2}-\frac{\sqrt{39}}{16}\leq \alpha ^{2}\leq \frac{1}{2}+\frac{\sqrt{39}}{16}
\end{equation}

Considering the potential applicability of copying quantum inseparability
in the field of quantum communications the range of \( \alpha ^{2} \) becomes crucial
since it defines which entangled states are accessible for broadcasting.
\\
This paper is organised as follows. First we use universal quantum
cloners\footnote{
The quantum cloners that we refer are \( 1\rightarrow 2 \) type unless otherwise stated
( Sec. 3).
} (in general less optimal) for local copying of the qubits to obtain
the nonlocal output state as a function of cloning machine reduction
factor and \( \alpha  \). The range of \( \alpha ^{2} \) is then provided as a function of the
reduction factor from which it follows that the range defined by (2)
is indeed the largest possible attainable using only local operations.
We also show that only those universal quantum cloners for which the
fidelity exceeds some threshold value are useful in the sense that
the nonlocal output state becomes inseparable for some values of \( \alpha  \).
Lastly we investigate the possibilty of broadcasting entanglement
into three pairs using only local operations.

\section{\large Broadcasting of entanglement using universal quantum cloner
for local copying}

Possibly the most prominent feature that distinguishes between classical
and quantum information theory is the ``no cloning theorem'' [2] which
prevents in producing perfect copies of an arbitrary quantum mechanical
state. The question `` how well then one can replicate a quantum mechanical
state ? '' lead to the concept of quantum cloning machine or simply
quantum cloner. Buzek and Hillary [3] were the first to construct
an universal quantum cloner which was later shown to be optimal by
Bru\( \ss  \) et al [4]. A \( 1\rightarrow 2 \) universal quantum cloner is defined as a quantum
mechanical device which takes a given qubit together with a blank
one as input to produce two qubits at the output. If the given unknown
input state is \( \left| \psi \right\rangle  \)(or , \( \rho _{in}=\left| \psi \right\rangle  \)\( \left\langle \psi \right|  \)) then each output state is given by the reduced
density operator \( \rho _{out}=\eta \left| \psi \right\rangle \left\langle \psi \right| +\frac{1}{2}\left( 1-\eta \right) I \). Since one can always write the density operator
of the input state in the form \( \left| \psi \right\rangle  \)\( \left\langle \psi \right|  \)\( =\frac{1}{2}(I+\overrightarrow{s}.\overrightarrow{\sigma }) \), therefore the output state can
also be written as \( \rho _{out}=\frac{1}{2}\left( I+\eta \overrightarrow{s}.\overrightarrow{\sigma }\right)  \), where \( I \) is the \( 2\times 2 \) identity matrix \( \overrightarrow{\sigma } \) represents
the set of Pauli spin matrices and \( \eta  \) is the reduction factor of the
original Bloch vector \( \overrightarrow{s} \). The measure of the quality of clones is defined
by the fidelity \( F=\left\langle \psi \right| \rho \left| \psi \right\rangle =\frac{1}{2}\left( 1+\eta \right)  \). Its clear that the maximal value of \( \eta  \) corresponds
to the optimal quantum cloner which therefore produces the best possible
replicas. The symmetry and isotropy conditions than an universal quantum
cloner satisfies have been discussed in details in Ref. [4]. \\
We first use the following universal cloning transformation for local
copying to broadcast entanglement. This is the simplest less optimal
cloning transformation requiring two more ancilla qubits, defined
by

\begin{equation}
\label{3}
U\left| 0\right\rangle \left| \right\rangle \left| Q\right\rangle =a\left| 00\right\rangle \left| A\right\rangle +b(\left| 01\right\rangle +\left| 10\right\rangle )\left| B\right\rangle 
\end{equation}

\begin{equation}
\label{4}
U\left| 1\right\rangle \left| \right\rangle \left| Q\right\rangle =a\left| 11\right\rangle \widetilde{\left| A\right\rangle }+b(\left| 01\right\rangle +\left| 10\right\rangle )\widetilde{\left| B\right\rangle }
\end{equation}

where \( \left| \right\rangle  \)denotes the blank qubit supplied to the cloner, \( \left| Q\right\rangle  \) denotes the
initial state of the quantum copier (ancilla), \( \left| A\right\rangle ,\left| B\right\rangle ,\widetilde{\left| A\right\rangle },\widetilde{\left| B\right\rangle } \) are the normalized
ancilla output states. The coefficients \( a \) and \( b \) are in general complex.
The following conditions hold from unitarity, isotropy and symmetry
requirements for an universal quantum cloner [4]

\begin{equation}
\label{5}
\left| a\right| ^{2}+2\left| b\right| ^{2}=1
\end{equation}

\begin{equation}
\label{6}
\left\langle B\right| \widetilde{\left. B\right\rangle }=\left\langle A\mid B\right\rangle =\widetilde{\left\langle A\right. }\widetilde{\left| B\right\rangle }=0
\end{equation}

The fidelity of the above universal quantum cloner defined by the
transformations \( (3) \) and \( (4) \) along with the conditions \( (5) \) and \( (6) \), is given
by

\begin{equation}
\label{7}
F=\frac{1}{2}(1+\eta )
\end{equation}

where the reduction factor \( \eta  \) is given by

\begin{equation}
\label{8}
\eta =\left| a\right| ^{2}=Re\left( b^{*}a\widetilde{\left\langle B\right| }\left. A\right\rangle +a^{*}b\widetilde{\left\langle A\right| }\left. B\right\rangle \right) 
\end{equation}

Choosing, \( \widetilde{\left\langle B\right. }\left| A\right\rangle =\widetilde{\left\langle A\right| }\left. B\right\rangle =1 \) one obtains the optimal quantum cloner [4] for which \( \eta =2/3 \).
Thus a less optimal quantum cloner but nevertheless universal (isotropic)
can be constructed by varying the scalar product of the ancilla output
states. 

Now consider two distant parties \( a_{1} \)and \( a_{2} \) share a pair of particles
prepared in the state

\begin{equation}
\label{9}
\left| \Psi \right\rangle =\alpha \left| 00\right\rangle +\beta \left| 11\right\rangle 
\end{equation}

where \( \alpha ,\beta  \) are real and \( \alpha ^{2}+\beta ^{2}=1. \) The first qubit belongs to \( a_{1} \) and the second
qubit belongs to \( a_{2} \). Now the two systems \( a_{i}(i=1,2) \) are locally copied according
to the cloning transformations (3) and (4) to produce output two systems
\( b_{i}(i=1,2) \). The local output state of a copier is given by the density operator

\footnote{
\( \left| +\right\rangle =\frac{1}{\sqrt{2}}\left( \left| 01\right\rangle +\left| 10\right\rangle \right)  \)
}
\begin{equation}
\label{10}
\widehat{\rho }_{a_{i}b_{i}}^{(out)}=\alpha ^{2}\eta \left| 00\right\rangle \left\langle 00\right| +\beta ^{2}\eta \mid \left| 11\right\rangle \left\langle 11\right| +(1-\eta )\left| +\right\rangle \left\langle +\right| \; \; 
\end{equation}

The nonlocal output is described by the density operator

\[
\widehat{\rho }_{a_{i}b_{j}}^{(out)}=[\alpha ^{2}\eta +(\frac{1-\eta }{2})^{2}]\left| 00\right\rangle \left\langle 00\right| +[\beta ^{2}\eta +(\frac{1-\eta }{2})^{2}]\left| 11\right\rangle \left\langle 11\right| \]

\begin{equation}
\label{11}
+(\frac{1-\eta ^{2}}{4})(\left| 01\right\rangle \left\langle 01\right| +\left| 10\right\rangle \left\langle 10\right| )+\alpha \beta \eta ^{2}(\left| 00\right\rangle \left\langle 11\right| +\left| 11\right\rangle \left\langle 00\right| )\; \; \; i\neq j
\end{equation}

It follows from the Peres-Horodecki theorem [7,8] that \( \widehat{\rho }_{a_{i}b_{j}}^{(out)} \) is inseparable
if

\begin{equation}
\label{12}
\frac{1}{2}-[\frac{1}{4}-\frac{(1-\eta ^{2})^{2}}{16\eta ^{4}}]^{1/2}\leq \alpha ^{2}\leq \frac{1}{2}+[\frac{1}{4}-\frac{(1-\eta ^{2})^{2}}{16\eta ^{4}}]^{1/2}
\end{equation}

The requirement that \( \left[ \frac{1}{4}-\frac{\left( 1-\eta ^{2}\right) ^{2}}{16\eta ^{4}}\right]  \) has to be positive otherwise the domain of
\( \alpha ^{2} \) would be meaningless leads to the lower bound of \( \eta  \),

\begin{equation}
\label{13}
\eta \geq \sqrt{\frac{1}{3}}
\end{equation}

The upper bound is of course \( 2/3 \) corresponding to the optimal quantum
cloner. 

Again applying the Peres-Horodecki theorem it is easy to obtain that
\( \widehat{\rho }_{a_{i}b_{i}}^{(out)} \) is separable if

\begin{equation}
\label{14}
\frac{1}{2}-\{\frac{1}{4}-\frac{(1-\eta )^{2}}{4\eta ^{2}}\}^{1/2}\leq \alpha ^{2}\leq \frac{1}{2}+\{\frac{1}{4}-\frac{(1-\eta )^{2}}{4\eta ^{2}}\}^{1/2}
\end{equation}

As one can observe comparing (12) and (14) that \( \widehat{\rho }_{a_{i}b_{i}}^{(out)} \) is separable if
\( \widehat{\rho }_{a_{i}b_{j}}^{(out)} \) is inseparable. \\
The range of \( \alpha ^{2} \) defined by (12) is a decreasing function of \( \eta  \), maximum
for \( \eta =2/3 \)(as given by (2)) and reduces to point set for \( \eta =1/\sqrt{3} \). For \( \eta =1/\sqrt{3} \) the only
entangled state for which broadcasting is possible is the maximally
entangled one. Thus the choice of an optimal quantum cloner is the
best for local cloning of the individual subsystems because maximum
number of entangled states are then available for broadcasting. We
also note from (13) that not all universal quantum cloners are suitable
for local cloning of entanglement. Only those universal quantum cloners
whose fidelity is greater than \( \frac{1}{2}(1+\sqrt{\frac{1}{3}}) \) are suitable because only then the
nonlocal output states becomes inseparable for some values of \( \alpha  \). \\
Though it may seem that the range of \( \alpha ^{2} \) given by (2) is the largest
still the possibilty of obtaining a larger range is not excluded because
we haven't so far considered the use of most general universal cloning
transformation to carry out local copying. \\
We now consider the following most general universal cloning transformation
satisfying the symmetry and isotropy requirements. The coefficients
are in general complex. The transformation is defined as,

\begin{equation}
\label{15}
U\left| 0\right\rangle \left| \right\rangle \left| Q\right\rangle =a\left| 00\right\rangle \left| A\right\rangle +b(\left| 01\right\rangle +\left| 10\right\rangle )\left| B\right\rangle +c\left| 11\right\rangle \left| C\right\rangle 
\end{equation}

\begin{equation}
\label{16}
U\left| 1\right\rangle \left| \right\rangle \left| Q\right\rangle =a\left| 11\right\rangle \widetilde{\left| A\right\rangle }+b(\left| 01\right\rangle +\left| 10\right\rangle )\widetilde{\left| B\right\rangle }+c\left| 00\right\rangle \widetilde{\left| C\right\rangle }
\end{equation}

along with the constraints following from unitarity, symmetry and
isotropy conditions [4],

\begin{equation}
\label{17}
\left| a\right| ^{2}+2\left| b\right| ^{2}+\left| c\right| ^{2}=1
\end{equation}

\begin{equation}
\label{18}
a^{*}c\left\langle A\right| \widetilde{\left. C\right\rangle }+2\left| b\right| ^{2}\left\langle B\right| \widetilde{\left. B\right\rangle }+ac^{*}\left\langle C\right. \widetilde{\left| A\right\rangle }=0
\end{equation}

\begin{equation}
\label{19}
\left| a\right| ^{2}-\left| c\right| ^{2}=Re\left( b^{*}a\widetilde{\left\langle B\right| }\left. A\right\rangle +a^{*}b\widetilde{\left\langle A\right| }\left. B\right\rangle \right) 
\end{equation}

\begin{equation}
\label{20}
Im\left( b^{*}a\widetilde{\left\langle B\right| }\left. A\right\rangle +a^{*}\widetilde{b\left\langle A\right| }\left. B\right\rangle \right) =0
\end{equation}

\begin{equation}
\label{21}
b^{*}c\left\langle B\right. \widetilde{\left| C\right\rangle }+c^{*}b\left\langle C\right. \widetilde{\left| B\right\rangle }=0
\end{equation}

\begin{equation}
\label{22}
ab^{*}\left\langle B\mid A\right\rangle +bc^{*}\left\langle C\mid B\right\rangle =0
\end{equation}

\begin{equation}
\label{23}
ab^{*}\widetilde{\left\langle B\right| }\widetilde{\left. A\right\rangle }+bc^{*}\widetilde{\left\langle C\right. }\widetilde{\left| B\right\rangle }=0
\end{equation}

\begin{equation}
\label{24}
c^{*}a\widetilde{\left\langle C\right. }\left| A\right\rangle =a^{*}c\widetilde{\left\langle A\right| }\left. C\right\rangle 
\end{equation}

The reduction factor \( \eta  \) of this universal quantum cloner is \( (\left| a\right| ^{2}-\left| c\right| ^{2}) \), given
by  (19). \\
A lengthy but straightforward calculation shows that when the above
defined universal quanum cloner is used for local copying to broadcast
entanglement the nonlocal output state obtained is the same as given
by (11). Hence we find that the nonlocal output density operator retaining
the same form even though the most general universal cloning transformation
is used for local cloning of the subsystems. What we mean is that
\( \eta  \) happens to be present in the density operator in the same way as
in (11). The only difference is that in the later case the reduction
factor is given by \( \left| a\right| ^{2}-\left| c\right| ^{2} \) whereas in the former one it is just \( \left| a\right| ^{2} \) but in
both cases, the reduction factor is a function of scalar products
of the ancilla output states. \\
That only those universal quantum cloners whose reduction factor is
greater than or equal to \( \frac{1}{\sqrt{3}} \) can be used for local copying for the purpose
of broadcasting entanglement is evident when we write the output density
operator (11) in the scaled form. Note that iff the original entangled
state is maximally entangled then only the output state can be expressed
in a scaled form. Thus for \( \alpha =\beta =\frac{1}{\sqrt{2}} \) the output state density operator can
be expressed as,

\begin{equation}
\label{25}
\widehat{\rho }^{out}=s\left| \psi \right\rangle \left\langle \psi \right| +(\frac{1-s}{4})I
\end{equation}

where \( s=\eta ^{2} \). \\
\( s \) is the scaling parameter which goes as square of the reduction factor.
We know that Werner states have the same form as (25) and are separable
if \( s<1/3 \)[8]. Thus the output state described by (25) is separable when
\( \eta <1/\sqrt{3} \) \( \left( s=\eta ^{2}\right)  \). \\
From inequality (14) it is obvious that applying local cloning on
subsystems cannot broadcast a pure entangled state if it cannot do
the same for maximally entangled states. Thus we can conclude that
when \( \eta <1/\sqrt{3} \) no broadcasting is possible for any pure entangled input state.

\section{\large Broadcasting entanglement into three pairs}

We have seen that its possible in general to broadcast entanglement
into two pairs and an optimal broadcasting results by using an optimal
quantum cloner to carry out local copying. So the next question is
whether one can optimally broadcast the original entanglement shared
by a single pair into more than two pairs. The simplest possible case
is the \( 1\rightarrow 3 \) entanglement broadcasting, which we consider here. The procedure
is essentially same as the \( 1\rightarrow 2 \) case. Only in this case one has to use
an optimal \( 1\rightarrow 3 \) quantum cloner [5] for local copying. \\
The necessary cloning transformation is defined as follows [5],

if the initial state to be cloned is \( \left| \phi \right\rangle =\alpha \left| 0\right\rangle +\beta \left| 1\right\rangle  \), we have

\begin{equation}
\label{26}
U(\left| \phi \right\rangle \otimes \left| \right\rangle \left| \right\rangle )=\alpha \left| \phi _{1}\right\rangle +\beta \left| \phi _{2}\right\rangle 
\end{equation}

where

\begin{equation}
\label{27}
\left| \phi _{1}\right\rangle =U\left| 0\right\rangle \left| \right\rangle \left| \right\rangle =\sum ^{i=2}_{i=0}a_{i}\left| A_{i}\right\rangle \otimes \left| \{0,3-i\},\{1,i\}\right\rangle 
\end{equation}

\begin{equation}
\label{28}
\left| \phi _{2}\right\rangle =U\left| 1\right\rangle \left| \right\rangle \left| \right\rangle =\sum ^{i=2}_{i=0}a_{i}\left| A_{2-i}\right\rangle \otimes \left| \{0,i\},\{1,3-i\}\right\rangle 
\end{equation}

and \( a_{i}=\sqrt{\frac{3-i}{6}} \). \\
Here \( \left| A_{i}\right\rangle  \) are the orthogonal normalized output states of the ancilla
and \( \left| \{0.3-i\},\{1,i\}\right\rangle  \) denotes the symmetric and normalized states of three qubits
where \( (3-i) \) of them are in state \( \left| 0\right\rangle  \)and \( i \) are in state \( \left| 1\right\rangle  \).

The original entangled state shared by a single pair is given by

\begin{equation}
\label{29}
\left| \psi \right\rangle _{a_{1}a_{2}}=\alpha \left| 00\right\rangle _{a_{1}a_{2}}+\beta \left| 11\right\rangle _{a_{1}a_{2}}
\end{equation}

We now apply this cloner for local copying each qubit \( a_{1} \) and \( a_{2} \) . Thus
we get a compound system consisting of six spin \( 1/2 \) particles. The objective
is to find whether the nonlocal output states are inseparable simultaneously
for some values of \( \alpha  \) for which the local output states are separable.
We first test the inseparability of the nonlocal output states. \\
We therefore write the nonlocal output state described by the density
operator

\[
\widehat{\rho }^{out}=(\frac{45\alpha ^{2}+4}{81})\left| 00\right\rangle \left\langle 00\right| +(\frac{45\beta ^{2}+4}{81})\left| 11\right\rangle \left\langle 11\right| \]
 
\begin{equation}
\label{30}
+\frac{14}{81}(\left| 01\right\rangle \left\langle 01\right| +\left| 10\right\rangle \left\langle 10\right| )+\frac{25\alpha \beta }{81}\left( \left| 00\right\rangle \left\langle 11\right| +\left| 11\right\rangle \left\langle 00\right| \right) 
\end{equation}

Applying the Peres-Horodecki theorem [7,8] we find that the above
state is separable. In fact this can also be seen if one writes the
density operator in the scaled form. For \( \alpha =\beta =1/\sqrt{2} \)

\begin{equation}
\label{31}
\widehat{\rho }^{out}=s\left| \psi \right\rangle \left\langle \psi \right| +(\frac{1-s}{4})I
\end{equation}

where \( s=\eta ^{2}=25/81 \), \( \eta =5/9 \) is the reduction factor corresponding to the \( 1\rightarrow 3 \) optimal
quantum cloner defined by the transformations (19) and (20). Separability
requires that \( s<1/3 \) which is satisfied as can be easily seen.

\section{\large Comparison with nonlocal cloning of entanglement}

Recently it has been shown that quantum inseparability can be copied
better (in the sense much larger range of \( \alpha  \) can be achieved) by using
a nonlocal copier [9] than when two local copiers are used [1]. The
range of \( \alpha ^{2} \) in the case of nonlocal cloning of entanglement is given
by [9]

\begin{equation}
\label{32}
\frac{1}{2}-\frac{\sqrt{2}}{3}\leq \alpha ^{2}\leq \frac{1}{2}+\frac{\sqrt{2}}{3}
\end{equation}

which is much wider than the range given by (2). \\
Here it is also worth comparing local cloning and nonlocal cloning
of entanglement. It will be interesting to see maximum how many copies
of entanglement can be made by nonlocal cloning. For \( 1\rightarrow M \) nonlocal cloning
of entanglement the output that can always be written in a scaled
form [10] is given by

\begin{equation}
\label{33}
\widehat{\rho }^{out}=s_{nl}\left| \psi \right\rangle \left\langle \psi \right| +\left( \frac{1-s_{nl}}{4}\right) I
\end{equation}
 ( subscript \( nl \) stands for nonlocal)

where the scaling parameter \( s_{nl}=\frac{4+M}{5M} \), \( M \) being the number of copies of entanglement.
For the output state to be separable for all entangled pure states
\( \left| \psi \right\rangle  \), we require that \( s_{nl}<1/3 \) which is satisfied for \( M=7 \). Thus a nonlocal cloning
of entanglement despite being difficult to implement in practice can
produce a maximum of six copies of entanglement whereas local cloning
of entanglement can produce only two. Intuitively one can also understand
the above result. We know that local operations (if not unitary) inevitably
results in loss of entanglement but there are no such restrictions
on nonlocal operations. What happens in nonlocal case is that the
entanglement of the system is actually being copied. But in local
cloning of entanglement cloning operations are applied on the individual
subsystems and entanglement of the nonlocal output comes as a biproduct.
Therefore, in nonlocal cloning, the bipartite system as a whole gets
entangled with a single cloning machine , whereas in local cloning
\em each \em individual subsystem separately gets entangled with a
cloning machine. Thus the entanglement transfer to the machine is
larger in the local cloning case. So its not surprising that nonlocal
cloning will produce more copies of entanglement than the local cloning.

\section{\large Conclusion}

We have discussed the role of an universal quantum cloner (in general
less optimal) used for local copying the subsystems in broadcasting
of entanglement. In particular we have shown that quantum inseparability
is best copied when one uses optimal quantum cloner. We also pointed
out that only those universal quantum cloners are useful for local
copying  whose fidelity exceeds a threshold value to broadcast entanglement.
We also showed that broadcasting of entanglement into more than two
pairs is forbidden using only local operations.

\( \medskip  \)

{\bfseries \large Acknowledgements\par}

We thank Dagmar Bru\( \ss  \) for pointing out an error in our earlier manuscript
and helpful discussions. One of the authors S.B. is grateful to Dipankar
Home for introducing him into this field.

\( \medskip  \)

{\bfseries \large References\par}

[1] V. Buzek, V. Vedral, M.B. Plenio, P.L. Knight, and M. Hillery,
Phys. Rev. A 55 (1997) 3327.

[2] W. K. Wootters and W. H. Zurek. Nature (London) 299 (1982) 802.

[3] V. Buzek and M. Hillary, Phys. Rev. A 54 (1996) 1844.

[4] D. Bru\( \ss  \), D. P. DiVincenzo, A. Ekert, C. A. Fuchs, C. Macchiavello
and J. A. Smolin, Phys. Rev. A 57 (1998) 2368.

[5] N. Gisin and S. Massar, Phys. Rev. Lett. 79 (1997) 2153. 

[6] D. Bru\( \ss  \), A. Ekert, and C. Macchiavello, quant-ph/9712019.

[7] A. Peres, Phys. Rev. Lett. 77 (1996) 1413.

[8] M. Horodecki, P. Horodecki, and R. Horodecki, Phys. Lett. A 223
(1996) 1.

[9] V. Buzek and M. Hillary, Phys. Rev. Lett. 81 (1998) 5003.

[10] R. F. Werner, Phys. Rev. A 58 (1998) 1827.

\end{document}